\documentstyle[11pt,aaspp]{article}
\def\be{\begin{equation}}
\def\ee{\end{equation}}

\def\refitem #1! #2! #3! #4;{\hang\noindent
    \hangindent 20pt #1, #2, #3, #4.\par}

\def\vs     {{\it vs.}}
\def\etal   {{\sl et~al.}}

\def\wisk#1{\ifmmode{#1}\else{$#1$}\fi}

\def\lsim   {\wisk{_<\atop^{\sim}}}
\def\gsim   {\wisk{_>\atop^{\sim}}}

\begin{document}

\noindent
\makebox[0pt][l]{
\raisebox{36pt}[0pt][0pt]{
\hskip 3in
astro-ph/9410070, UCLA-ASTRO-ELW-94-04}}

\title{Comments on the Quasi-Steady-State Cosmology}

\author{Edward L. Wright}
\affil{UCLA Astronomy, Los Angeles CA 90024-1562}

\begin{abstract}
The Quasi-Steady-State Cosmology as proposed by Hoyle, Burbidge
and Narlikar does not fit the observed facts of the Universe.
In particular, it predicts that 75-90\% of the radio sources
in the brightest sample that shows steeper than Euclidean
source counts should be blueshifted.
\end{abstract}

\section{Introduction}

The quasi-steady-state cosmology (QSSC)
of Hoyle, Burbidge \& Narlikar (1994, hereafter
HBN) is based on a spatially flat Universe with a scale factor
having an exponential growth multiplying a sinusoidal oscillation.
Since the observed expansion of the Universe is primarily due to the
oscillation, this QSSC model provides 200 Gyr in which to produce the
cosmic microwave background (CMB), 
instead of the $0.25/H_\circ \approx 4$~Gyr
allowed in the Steady-State cosmology.  In addition, the large volume
of the Universe during its oscillatory maxima provides a mechanism for
steeper than Euclidean source counts.  In this paper I show that
it is still difficult, if not impossible, to produce the CMB in the
QSSC model proposed by HBN; and that producing the CMB is incompatible
with HBN's model for the faint blue galaxies.  The large millimeter
wave optical depth required to blacken the CMB makes the millimeter
wave luminosity of the QSO BR~1202-0725 the highest known luminosity.
I further show that
steeper than Euclidean source counts are always accompanied by a
redshift distribution that is dominated by blueshifts.

\section{The CMB Power Problem}

In the QSSC model, the CMB is produced by dust absorbing the
diffuse extragalactic background light (EBL), and re-emitting the energy
in the millimeter region.  The large millimeter opacity
required is produced by iron whiskers.  There is a relation
between the ratio of EBL to CMB energy densities and the visible
optical depth through an oscillatory minimum.
In this section I will consider a simple model, with a single
frequency bin for the ``optical'' light that heats the dust
which makes the CMB, and a second frequency bin for the CMB.
If I assume a grey opacity in the optical, then
the ``optical'' energy density $U_{EBL}$ satisfies the equation
\be
d[a^4 U_{EBL}]/dt = 4\pi a^4 j - \kappa c a^4 U_{EBL}
\ee
where $4\pi j$ is the luminosity density, $\kappa$ is the optical
depth per unit length, and $a$ is the scale factor of the Universe.
In the QSSC model 
$u$, $j$ and $\kappa$ are periodic functions of time with period
$Q$.
The scale factor follows
\be
a(t) \propto (1+\alpha \cos(2\pi t/Q)) \exp(t/P) = A(t) \exp(t/P)
\ee
where
HBN gives values 
$\alpha = 0.75$, $Q = 40$~Gyr, $P/Q = 20$, and $t_\circ/Q = 0.85$.
I will now assume that the luminosity density and dust density
follow
$
j = j_1/A^3
$
and
$
\kappa = \kappa_1/A^3.
$
With these assumptions it is easy to find periodic solutions for $U_{EBL}(t)$.
Define
\be
F(t) = \exp(-\kappa_1 c \int A^{-3}(t) dt)
\ee
and then
\be
U_{EBL} = a^{-4} F(t) \left( C_1 \int F^{-1}(t^\prime) 
    A(t^\prime)\exp(4t^\prime/P) dt^\prime  + C_0 \right).
\ee
The ratio of $C_1$ to $C_0$ can be adjusted to give a periodic 
solution.

The ``CMB'' energy density $U_{CMB}$ satisfies the equation
\be
d[a^4 U_{CMB}]/dt = + \kappa c a^4 U_{EBL}
\ee
where $\kappa$ is again the visual opacity.  Here I assume that any CMB 
radiation absorbed by dust is reradiated as CMB radiation, and the net
effect on the energy density cancels out.  The solution to this equation
is
\be
U_{CMB} = a^{-4} \left( \int \kappa_1 c A(t^\prime)\exp(4t^\prime/P) 
          U_{EBL} dt^\prime + U_0 \right)
\ee
and I choose the constant of integration $U_0$ to make $U_{CMB}$ a periodic 
function.

\begin{figure}[t]
\plotone{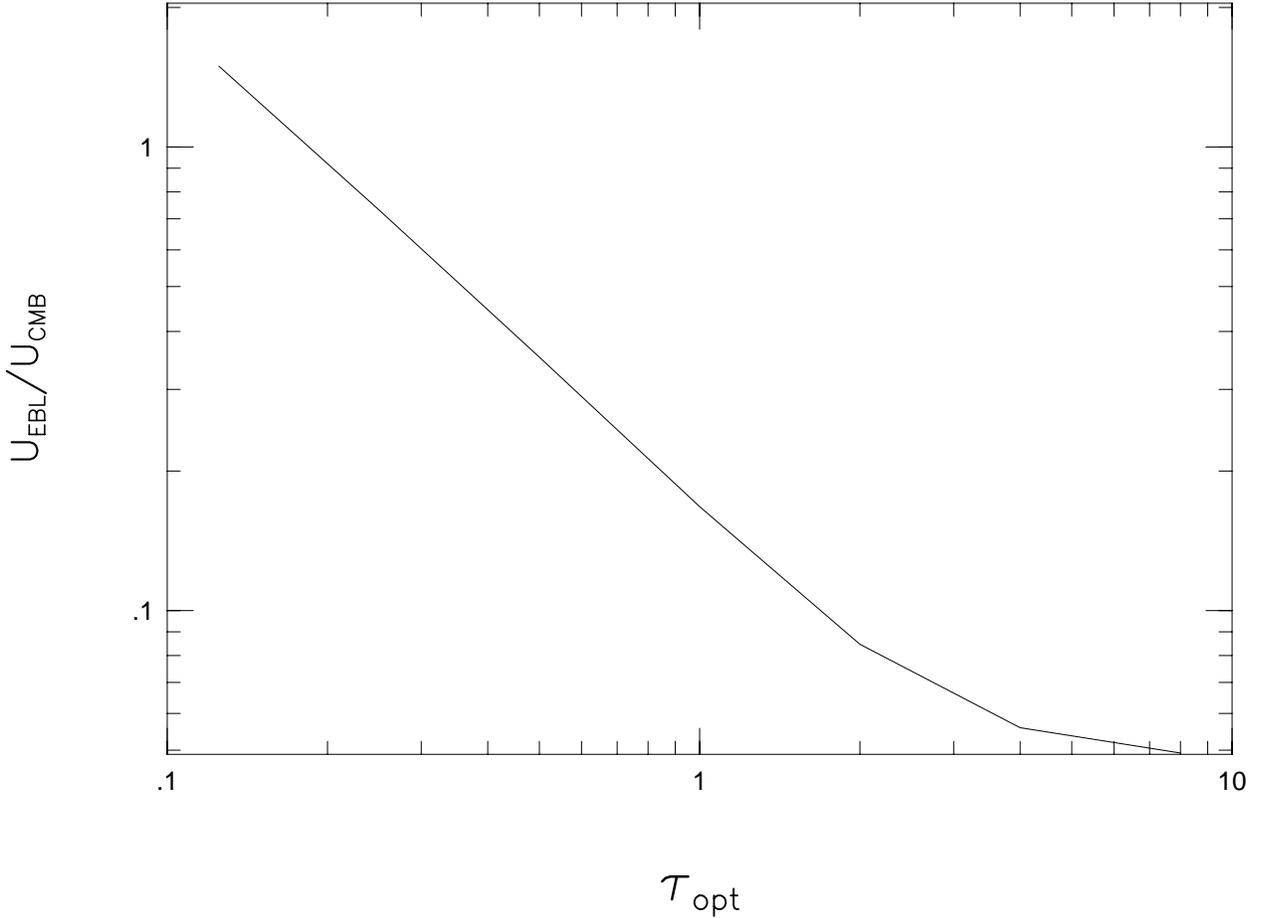}
\caption{Optical to CMB energy density ratio for different visual optical
depths through the oscillatory minimum.}
\label{ratvtau}
\end{figure}

The ratio of the optical light now (at $t_\circ/Q = 0.85$) to the
CMB now is shown as a function of the total visible optical depth through
an oscillation in Figure \ref{ratvtau}.  Even for $\tau_{opt} >> 1$
the required optical power remains higher than the observed limits on
the EBL.  An EBL brightness of 1 $S_{10}$ in the visible gives 
$U_{EBL}/U_{CMB} \lsim 0.01$, while the QSSC requires 
$U_{EBL}/U_{CMB} \geq 0.05$ even for $\tau_{opt} = 8$.

\begin{figure}[t]
\plotone{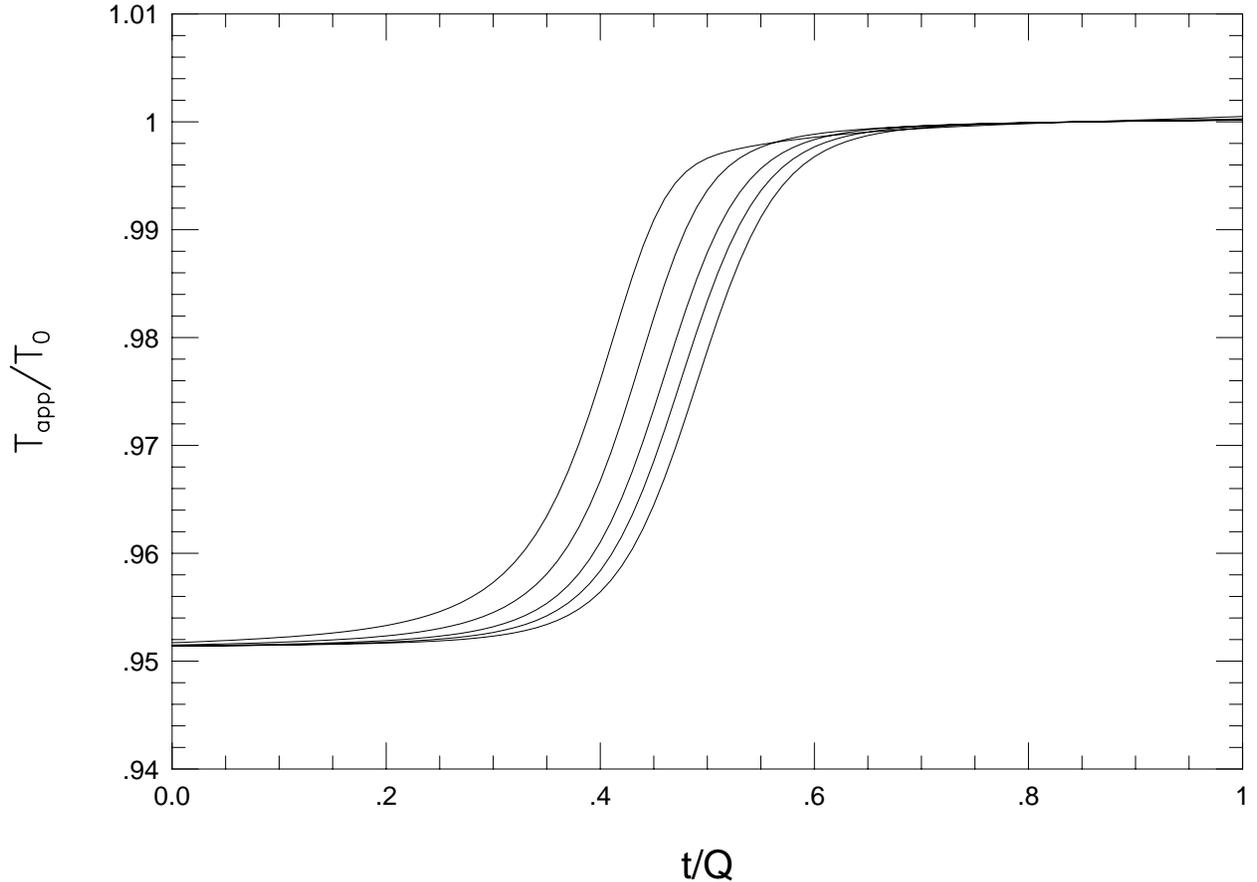}
\caption{Apparent dust temperature vs time.}
\label{tvst}
\end{figure}

The quantity $U_{CMB}+\tau_{opt}U_{EBL}/\tau_{mm}$ 
is $\propto T_{eq}^4$, where $T_{eq}$ is the 
equilibrium temperature of the dust, and $\tau_{mm}$ is the millimeter wave
optical depth through the oscillatory cycle.
But dust observed at a redshift
$1+z = 1/a$ has an apparent temperature of $T_{ap} = T/(1+z)$, which
is $\propto ((U_{CMB}+\tau_{opt}U_{EBL}/\tau_{mm})a^4)^{0.25}$.
This quantity is plotted versus
$t/Q$ for several values of $\tau_{opt}$ while letting
$\tau_{mm} \rightarrow \infty$ in Figure \ref{tvst}.
For larger values of $\tau_{opt}$ the 5\% jump in $T_{ap}$ slightly
before the oscillatory minimum instead of at the minimum.  This
slightly reduces the millimeter optical depth required to adequately
blacken the CMB.

For a grey opacity in the millimeter range, it is easy to show that
the required opacity is $\tau \gsim 4$ between now and the first 5\%
temperature step.  The spectrum is approximately a superposition 
of blackbodies:
$T_\circ$ with weight $1-e^{-\tau}$ and $0.95 T_\circ$ with
weight $e^{-\tau}$.  The resulting ${\rm var}(T)/T^2 = e^{-\tau}/20^2$.
But since ${\rm var}(T)/T^2 = 2y < 5 \times 10^{-5}$ according to
Mather {\it et al.} (1994), one needs $e^{-\tau} < 2y(P/Q)^2 = 0.02$,
or $\tau > 4$ to the oscillatory minimum.  Numerical integration of
${\rm var}(T)/T^2$ for a grey millimeter opacity shows that
$\tau_{mm} \geq 6$ gives a sufficiently small $y$ if $\tau_{opt}$ is
small, but $\tau_{mm} \geq 4$ suffices if $\tau_{opt} >> 1$.
The opacity curve given in HBN's Figure 5 is far from grey, but
this calculation agrees well with the HBN estimate of $\tau \approx 10$ 
through the minimum for adequate blackening of the CMB.

\section{Millimeter Point Sources at High Redshift}

The observations of McMahon {\it et al.} (1994)
essentially rule out the published version of the QSSC, since they
have observed millimetric fluxes from quasars with $z > 4$.
As discussed above, the QSSC requires that the millimeter
transmission between $t_\circ$ and the oscillatory minimum be
very close to zero, and these high redshift quasars are at or
beyond the oscillatory minimum.
The ratio of the optical depth to $z = 4.69$ for an observed
wavelength of 1.25 mm to the 1.4~GHz optical depth through
a full oscillation is 63.
This calculation is based on the opacity \vs\ $\lambda$ from
Figure 5 of HBN.
Since the 1.4~GHz optical depth should
be $> 0.125$ to give enough CMB flux at 1.4~GHz,
the quasars observed by McMahon {\it et al.} must
have gone through $\tau > 8$.  Thus the QSSC requires that these
sources be $>2000$ times more luminous in millimeter waves
than one would normally assume.  This raises $\nu L_\nu$ for
BR~1202-0725 to $5 \times 10^{15}\;L_\odot$.  If 3C273 had this
$\nu L_\nu$ in the V band its magnitude would be $V = 4.5$.

\section{Blueshifts}

\begin{figure}[t]
\plotone{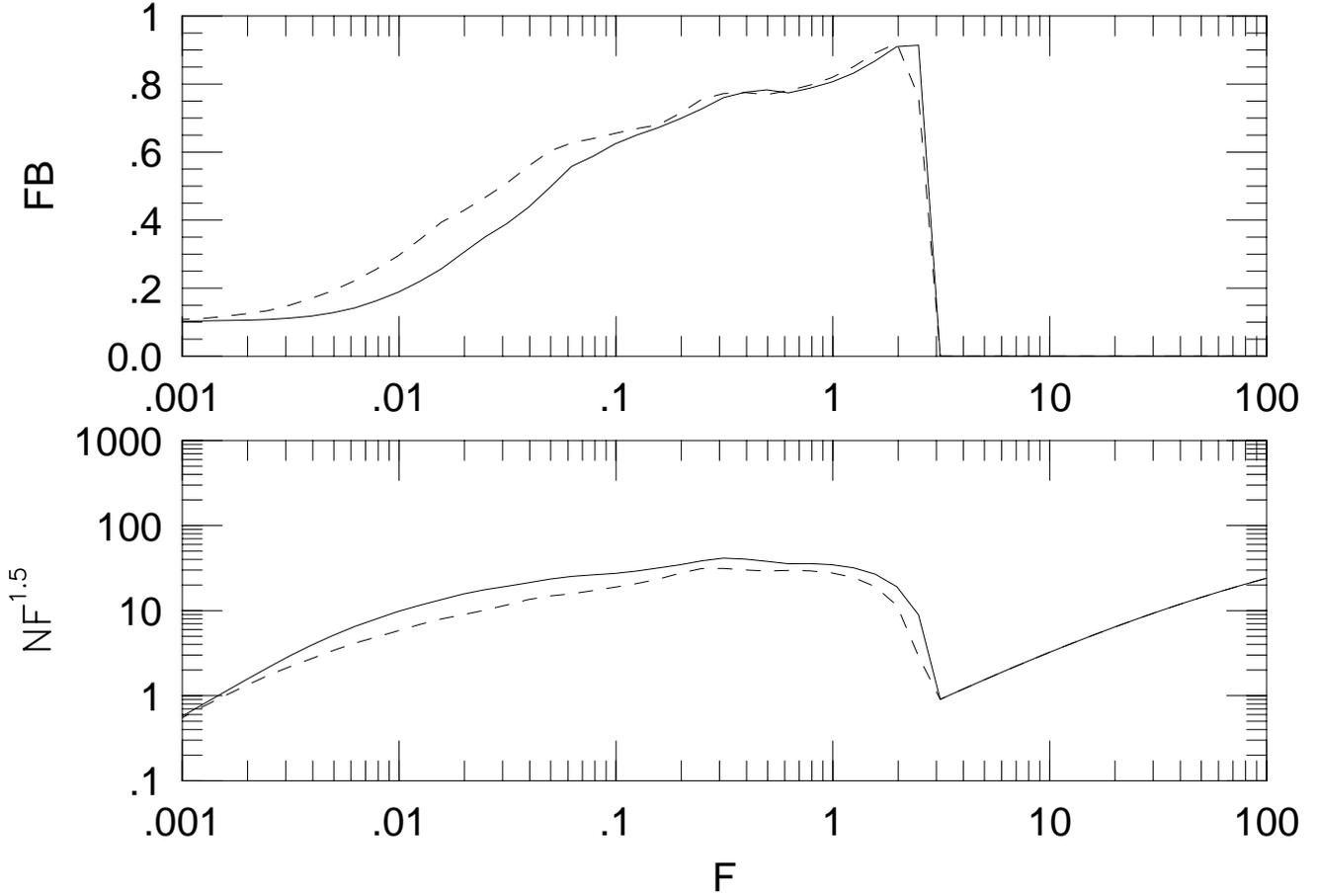}
\caption{Source counts and blueshifted fraction for the power law 
luminosity function used by HBN.  The dashed curves show the effect
of radio absorption with $\tau = 0.125$ at 21 cm.}
\label{hoylerf3}
\end{figure}

I have computed source counts using the
formulae given by HBN, which are the standard equations of relativistic
cosmology once the non-standard quasi-periodic $a(t)$ is assumed.
HBN have assumed that radio sources are uniformly distributed in
{\em physical} volume, rather than the usual assumption of a constant
comoving density.  This assumption maximizes the steepness of the
source counts, which is needed to fit the observed data, but it also
increases the blueshifted fraction.
The relevant equations are:
\be
(1+z) = a(t_\circ)/a(t)
\ee
and the comoving radius of our past light cone is
\be
r = \int_t^{t_\circ} (1+z) c dt.
\ee
The angular size distance is $r_A = r/(1+z)$ while the luminosity
distance is $r_L = (1+z)r$.
A source with luminosity $L_\nu = L_\circ (\nu/\nu_\circ)^{-\gamma}$,
where $\nu_\circ$ is the observed frequency now, has a flux
\be
F(\nu_\circ) = \frac{(1+z)^{1-\gamma}L_\circ} {4\pi r_L^2}.
\ee
The source counts are given by
\be
n(F)  = \int_{-\infty}^{t_\circ} 
\phi(4\pi r_L^2 (1+z)^{\gamma-1} e^\tau F) \; 4\pi r_L^2
(1+z)^{\gamma-1} e^\tau \; r_A^2 c dt
\label{count}
\ee
where $n(F) dF$ is the source count per steradian in the flux
range $[F,F+dF]$, the luminosity function
$\phi(L_\circ)dL_\circ$ is the number density of sources with luminosities in
the range $[L_\circ,L_\circ+dL_\circ]$,
and the blueshifted source counts are
\be
n_B(F)  = \int_{-\infty}^{t_\circ}
P_B(z) \; \phi(4\pi r_L^2 (1+z)^{\gamma-1} e^\tau F) \; 4\pi r_L^2
(1+z)^{\gamma-1} e^\tau \; r_A^2 c dt
\label{blue}
\ee
where $P_B(z) = 1$ if $z < 0$ and zero otherwise.
Figure \ref{hoylerf3}
shows the normalized cumulative source counts 
$N(F) F^{1.5} = F^{1.5} \int_F^\infty n(F^\prime) dF^\prime$
and the blueshifted fraction
$FB = N(F)^{-1} \int_F^\infty n_B(F^\prime) dF^\prime$
for the luminosity function used by HBN: a one decade wide function
going as $L^{-2.1}$.  While HBN did not specify
a radio spectral index, this figure has been computed for sources
with a $L_\nu \propto \nu^{-0.75}$ spectrum.  The solid curves were
computed with $\tau = 0$, while the dashed curves in this
figure show the effect of radio absorption when the optical depth at
1.4~GHz is $\tau = 0.125$ through the first oscillatory minimum.
Note that the blueshifted fraction jumps to more than 90\%
for the brightest sources that show a steeper than Euclidean source count.
Thus the 3CR sample, which has source counts that are significantly
steeper than Euclidean, should be dominated by blueshifts if the QSSC
were correct.  While it is possible to have no blueshifts in the 3CR
flux range by a suitable choice of the luminosity function, one then
predicts source counts that are substantially less steep than Euclidean,
producing a $> 3\sigma$ discrepancy in source count slopes
between the model and the data (Jauncey, 1975).
At the peak of $N F^{1.5}$, the blueshifted fraction predicted by the
QSSC is $> 75\%$.
The 1 Jy sample of Allington-Smith \etal\ (1988) is at the 
peak of $N F^{1.5}$, has a secure redshift completeness $> 80\%$
(Rawlings \etal, private communication),
and no blueshifts.  Since one can not fit a $> 75\%$ blueshifted fraction
into a $< 20\%$ unidentified fraction, the QSSC prediction is false.

\begin{figure}[t]
\plotone{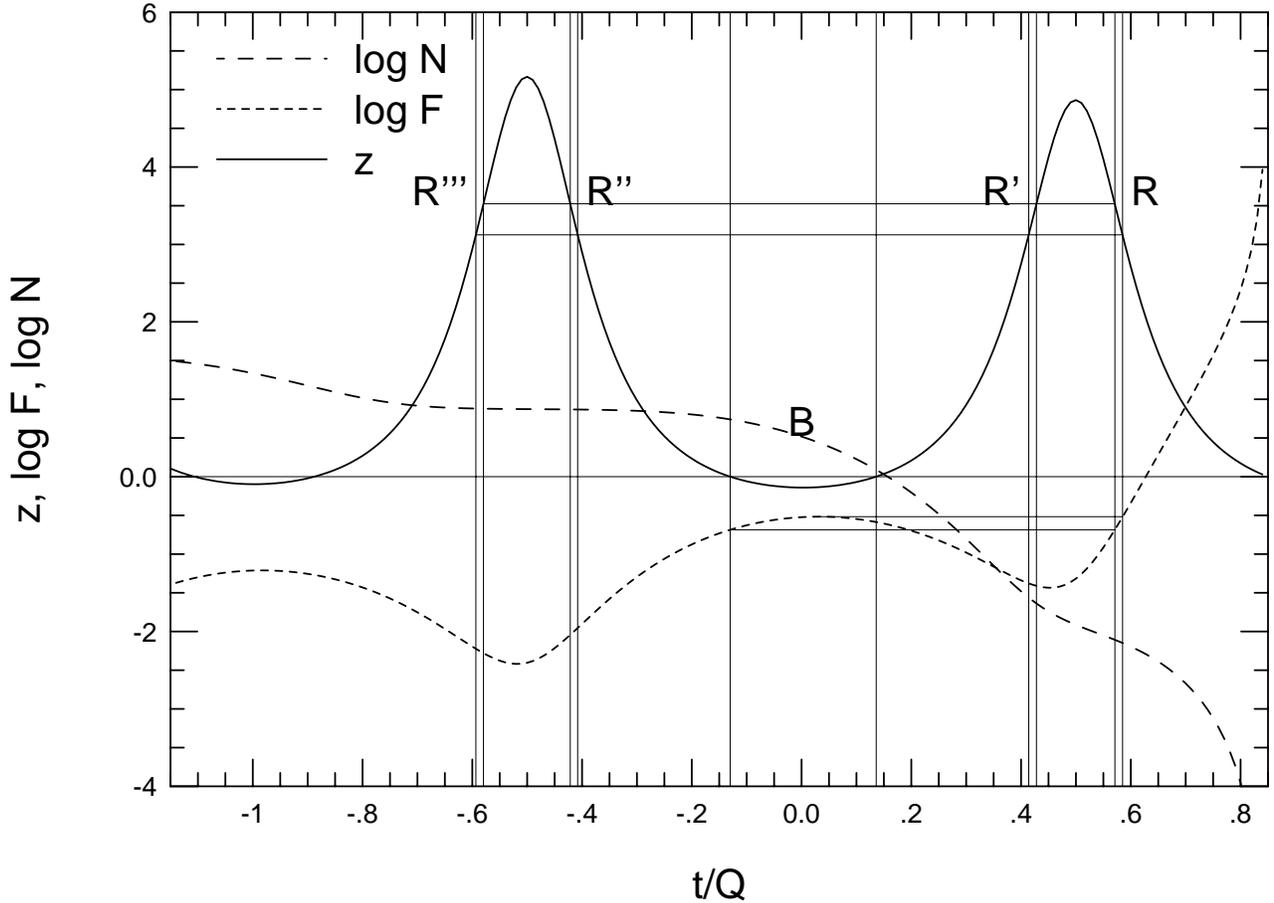}
\caption{The flux, redshift, and surveyed volume as function of emission
time t.}
\label{zfnvt}
\end{figure}

Another way of looking at this problem is shown in Figure \ref{zfnvt}.
In this plot the integrated physical volume is shown as $N$, with
\be
N = \int_t^{t_\circ} r_A^2 c dt.
\ee
The blueshifted region is labeled $B$.
The flux range in which blueshifted sources occur in the first oscillatory
maximum occur is shown by the lower two horizontal line segments, which
carry this flux range over to ``normal'' branch in the current half-cycle. 
In the region labeled $R$ the redshifted sources have the same fluxes
as the blueshifted sources.
The range of redshifts during this interval is $3.12 < z < 3.53$.
But the observed physical volume in the interval $R$ is 6000 times smaller
than the volume in the interval $B$.  Thus for every source in the interval
$R$ there should be 6000 blueshifted sources.  The flux intervals are 
identical, so this multitude of blueshifted sources should definitely be
present in radio surveys.  The 1 Jy sample has at least 1 source in the
interval $R$, a radio galaxy with redshift 3.4 (Lilly 1988), so it should
contain 6000 blueshifted sources if the QSSC were correct.
It is also possible that this galaxy is just past the oscillatory minimum,
in the region $R^\prime$.  But
this possibility only decreases the volume ratio to 1500, while adding
a factor of $\geq 5$ flux ratio.  Thus, if the galaxy were due to a source
in $R^\prime$, then
one would expect 1500 blueshifted sources in a sample of sources
brighter than 5 Jy at 408 MHz within the 1 Jy sample area of 0.11~sr.  
But this flux level corresponds to 9 Jy at 178 MHz for
$F_\nu \propto \nu^{-0.75}$, so all of these sources would be
in the 3C catalog.
But Spinrad \etal\ (1985) have redshifts for 214 out of the 235 3CR
sources with $|b| > 15^\circ$, a solid angle of $> 4$ sr, so this option
requires that 50,000 of the 21 3CR sources without measured redshifts
would have to be blueshifted.
Of course, the radio galaxy could be even more distant, in the
region marked $R^{\prime\prime}$.  But the blueshifted volume in $B$
is 250 times larger than the volume in $R^{\prime\prime}$ and the
flux ratio is $\geq 19$, so this explanation of the radio galaxy
requires 250 blueshifted sources brighter than 19 Jy at 408 MHz in
0.11~sr, or 25,000 such sources over the whole high galactic
latitude sky.
Finally, the region $R^{\prime\prime\prime}$ has a volume that
is 135 times smaller than the blueshifted volume in $B$, and a flux
ratio $\geq 34$.  This possibility requires 13,500 blueshifted
sources brighter than
34 Jy over the whole sky.  Thus for each possibility the QSSC
requires that there be many times more blueshifted sources than the
total number of observed sources.

I note that the Hewitt \& Burbidge QSO catalog (1993)
contains roughly 130 QSOs with $3.12 < z < 3.53$ but no blueshifts.

It is this sudden addition of the sources from the last oscillatory
maximum that creates the steep source counts in the QSSC model, but
the brightest of these sources, and hence the easiest to see, are
blueshifted.  The QSSC cannot have both steep source counts and no
blueshifts simultaneously, but the observed Universe does.  Hence the
QSSC is not a successful model for the Universe.

\section{Conclusion}

The QSSC is disproved by the same data that disproved the Steady State:
counts of bright radio sources.
This failure of the QSSC model, and the CMB power problem, were fully 
explained to both HBN and to the
editor of MNRAS in a referee's report dated July 1993, which included
a more extensive version of Figure \ref{hoylerf3}, but HBN 
proceeded to publish incorrect results.

The saddest part of this tale is that the QSSC can be compatible with
the radio source counts, the QSO $\langle V/V_{max} \rangle$ test, 
and the CMB;
by making $\alpha$ larger and $\tau$ large which eliminates the
blueshifts.  To get the steep source counts one adds a new factor
$[(1+\alpha \cos(2\pi t_\circ)/(1+\alpha \cos(2\pi t)]^n$ inside the
integrals in Equations \ref{count} and \ref{blue}.
This introduces density evolution into the QSSC, but in a periodic way.
While the philosophy of the Steady State model did not allow evolution
of the radio source population, the QSSC model certainly allows a
periodic evolution of the source population.
In fact, one might expect a high density of radio sources at the
oscillatory minima.
The only problem with this approach is that it eliminates the
observational differences between the QSSC and the Big Bang, and there
is nothing left to motivate the introduction of new physics.
But eliminating these differences is necessary, since the Big Bang can
fit the data, while HBN's QSSC model does not.

\end{document}